\documentclass[runningheads]{svmult}

\usepackage{epsfig}   
\usepackage{makeidx}   
\usepackage{graphicx}  
\usepackage{subeqnar}  
\usepackage{multicol}  
\usepackage{physmubb}  
\makeindex             

\newcommand{\sax}{{\it BeppoSAX}}
\begin{document}
\title*{X--Ray Observations of $\gamma$--Ray Burst Afterglows}
\toctitle{X--Ray Observations of $\gamma$--Ray Burst Afterglows}
\titlerunning{X--Ray Observations of $\gamma$--Ray Burst Afterglows}
%
\author{Filippo Frontera\inst{}}
\authorrunning{Filippo Frontera}
%
%
\institute{Physics Department, University of Ferrara, Ferrara, Italy \\
and \\
Istituto Astrofisica Spaziale e Fisica Cosmica, CNR, Bologna, Italy}

\maketitle              

\begin{abstract}
The discovery by the \emph{BeppoSAX} satellite of X-ray afterglow emission 
from the $\gamma$-ray burst which occurred on 28 February 1997 produced a revolution 
in our knowledge of the $\gamma$-ray burst phenomenon. Along with the discovery 
of X-ray afterglows, the optical afterglows of $\gamma$-ray bursts were 
discovered and the distance issue was settled, at least for long $\gamma$-ray bursts. 
The 30 year mystery of the $\gamma$-ray burst phenomenon is now on the way to solution. 
Here I rewiew the observational status of the X-ray afterglow emission, 
its mean properties (detection rate, continuum spectra, line features, 
and light curves), and the X-ray constraints on theoretical models of 
$\gamma$-ray bursters and their progenitors. I also discuss the early 
onset afterglow emission, the remaining questions, and the role of 
future X-ray afterglow observations.
\end{abstract}

\section{Introduction}
\label{intro}

The investigation into the nature and origin of celestial Gamma-Ray Bursts
(GRBs) has been one of the major challenges of high energy astrophysics,
since they were discovered by chance about 30 years ago with the \emph{VELA}
satellites \cite{Klebesadel73}.
The GRB origin remained
uncertain even when the \emph{Burst and Transient Source Experiment (BATSE)} 
experiment aboard the {\it Compton Gamma-Ray Observatory (CGRO)} in the 1990s 
provided a much larger
sample of GRBs (about 3000 events). Many outstanding results were obtained
with BATSE, among them  the completely isotropic distribution on the
sky of  GRBs and the inconsistency of the GRB number versus peak intensity
relation with a homogeneous volume distribution in space, inplying a deficit of faint bursts
\cite{Briggs96,Meegan92,Paciesas99}.
Two main explanations were given for these statistical properties: either
GRBs have their origin in neutron stars distributed in a large Galactic halo 
at more than 100 kpc distance
\cite{Brainerd92,Bulik98,Hakkila94,Lamb95,Podsiadlowski95,Shklovskii85} or
GRBs have an origin in sources at cosmological distances
\cite{Meegan92,Meszaros93a,Meszaros93b,Meszaros94,Narayan92,Paczynski90,Rees92,Shemi90}.

It was recognized (see, e.g., \cite{Fishman95}) that the identification of
a GRB counterpart at other wavelengths  was needed in order to make a 
breakthrough in
setting the distance scale of GRBs. Now the distance scale issue of at least the long ($>1$~s) 
GRBs  has been definitely settled thanks to the
X--ray astronomy mission \sax, an Italian satellite with
Dutch participation\cite{Boella97a}. \sax\  has also provided the most exciting 
results of the last six years in GRB astronomy. 
The high performance of \sax\ for GRB studies is due
to a particularly well-matched configuration of its payload, with both Wide Field
 and Narrow Field Instruments (WFIs and NFIs). The WFIs consist of a
$\gamma$--ray (40--700 keV) all--sky monitor, the \emph{Gamma-Ray Burst Monitor} 
(\emph{GRBM}) \cite{Frontera97}, and two Wide-Field Cameras (WFCs) instruments 
covering the 2--28 keV energy band \cite{Jager97}.
The NFIs include four focusing X--ray (0.1--10 keV) telescopes 
(one \emph{Low-Energy Concentrator Spectrometer (LECS)}\cite{Parmar97}
and three \emph{Medium-Energy Concentrator Spectrometers (MECS)}
\cite{Boella97b}). The NFIs also have two higher energy direct viewing
 detectors (the \emph{High Pressure Gas Scintillation Proportional Counter 
(HPGSPC)} \cite{Manzo97} and the \emph{Phoswich Detector System (PDS)}
 \cite{Frontera97}). 

The \sax\ capability of precisely localizing GRBs  was soon tested, only one
month after first light of the satellite, with the detection of
GRB960720 \cite{Piro98a}. Starting from December 1996, at the \sax\
Science Operation Center, an alert procedure for the search of simultaneous
\emph{GRBM} and \emph{WFCs} detections of GRB events was implemented. 
Thanks to this procedure, the first X--ray afterglow of a
GRB, GRB970228, was discovered on 28 February 1997 \cite{Costa97}. 
Simultaneously, an optical counterpart of the X--ray
afterglow source was detected \cite{Jvp97}. Less than two months later,
the first optical redshift ($z = 0.835$) of GRB970508 was obtained 
thanks to the prompt localization  provided by \sax, \cite{Metzger97}.
The location of GRBs at cosmic distances was definitely established.
At the same time, an exciting result on the GRB source
properties  came from radio observations of the same event: the
radio afterglow exhibited rapid variations for about one month due 
to interstellar scintillations \cite{Frail97}. 
Thus, only a very compact source, expanding highly relativistically, 
could be the origin of GRBs \cite{Frail97}.

Since the first detection of an X--ray afterglow, many other GRB afterglows
 have been detected, mainly with \sax\ and, less often, with
other satellites, i.e., the R\"{o}ntgensatellit ({\it ROSAT}), the {\it Rossi X--ray Time
Explorer (RXTE)}, the {\it Advanced Satellite for Cosmology and Astrophysics (ASCA)}, 
the {\it Chandra X--ray Observatory-AXAF (CHANDRA)}, and the
{\it XMM-Newton (XMM)}. In this paper I review the status of our knowledge of GRB
X--ray afterglows and discuss their implications.
In Sect.\ref{s:obs},  I discuss the measured afterglow properties
(detection rate, continuum spectral properties, light curves and X--ray line
features); in Sect.~\ref{s:early}, 
I discuss the early afterglow emission and its distinctive features with
respect to the prompt emission; and in Sect\ref{s:issues}, I summarize
the open questions and the prospects for future X--ray afterglow observations.

\section{X--Ray Afterglow Observations and Properties}
\label{s:obs}   
Tables~\ref{t:tab1}, \ref{t:tab2} and \ref{t:tab3} summarize the main results
of the X--ray searches for GRB afterglows. In the following provides further elaboration.

\subsection{X--Ray Afterglow Detection Rate}
\label{s:rate}
Table \ref{t:tab1} lists the information on GRB follow-up observations of X--ray 
afterglows  as of December 2001. Two
X--ray rich events, GRB011030 \cite{Gandolfi01} and GRB011130
\cite{Ricker01a}, were not included in the list.
GRB011030
was observed by {\it Chandra} \cite{Harrison01c} following the discovery
of a radio transient in the GRB error box \cite{Taylor01} and an X--ray source
was detected but no fading observed \cite{Harrison01c}. GRB011130 was also
observed by {\it Chandra} and several X--ray sources were detected
but no conclusions could be drawn 
as to their association with the GRB \cite{Ricker01b}.
As can be seen, most of the 43 GRBs listed in Table~\ref{t:tab1} were
promptly localized with the \sax\ WFCs, with the \emph{RXTE Proportional Counter Array (PCA)}, 
the \emph{All Sky Monitor (ASM)} and the 3rd \emph{Interplanetary Network (IPN)} 
contributing some additional examples.
The minimum starting time of the follow-up observations ranges from 2--3 hr
for \emph{RXTE}, to 5--6 hr for \sax, to longer times for the other
satellites. Given its limited sensitivity, the \emph{RXTE PCA} has provided
positive results only for very strong afterglow fluxes
(few mCrab) and for a limited time interval. Unfortunately most of
these PCA events were not followed up with more sensitive instruments. 

The criterion for
establishing the afterglow nature of an X--ray source is its fading
on time scales of a few hours or days. Unfortunately, for the weakest sources
this fading could not be well established, as the detected flux 
was very close to the instrument sensitivity limit. 
In these cases the source found in the GRB error box could be
a serendipitous field source unrelated to an X--ray afterglow, even though 
the probability of such a chance coincidence is low (see the notes to Table~\ref{t:tab1}).
With this caveat in mind, 
37 of the followed 43 GRBs (86\%), show X--ray
afterglows at a level $> 10^{-13}$~erg~cm$^{-2}$~s$^{-1}$ in the 2--10 keV band.
Thus, the detection rate of X--ray afterglows  is about
two times higher than that for optical afterglows (see the chapter by E. 
Pian).  
Therein lies the mystery of the so-called "dark bursts", i.e., bursts 
which show an X--ray  afterglow but not optical emission 
(see Section~\ref{s:issues}).

\subsection{X-Ray Afterglow  Continuum Spectra}
\label{s:spectrum}
The spectral shape of the X--ray afterglow emission gives important 
information about establishing the emission mechanisms. 
An initial question was whether the X--ray continuum emission 
was thermal or nonthermal.
The X--ray afterglow of GRB970228 showed \cite{Frontera98b} that, 
in the X-ray range (0.1--10 keV), a blackbody model was not acceptable.
The photon spectrum was well fit with a simple photoelectrically 
absorbed power-law (see Table~\ref{t:tab2}).
This spectral shape has been found in all X--ray afterglows known,
although for some GRBs (GRB980329 \cite{Zand98} and 
GRB981226 \cite{Frontera00a}), in which thermal models were tested, 
it was found that the data do not have sufficient signal-to-noise 
to discriminate between power--law and blackbody models.

Table~\ref{t:tab2} reports the best fit power--law photon indices \emph{I} 
(where the photon index \emph{I} is related to the flux density \emph{F} 
spectral index $\alpha$, $F_\nu \propto \nu^{+\alpha}$, by $I = \alpha -1$) and
absorbing hydrogen column densities $n_H$. The values reported in square
brackets were kept fixed during the fit. For comparison,  the
Galactic hydrogen column densities $n_{\rm H}^{\rm G}$  values 
along the GRB directions are  also reported \cite{Dickey90}. 
The best fit values of $I $ and $n_{\rm H}$ are available for 
only the brightest of the detected afterglows.
Most of the spectra have been determined up to 10 keV,
and in only one case (GRB990123) was the source detected up to 60 keV with the 
\sax\ PDS instrument, also confirming the power--law shape.

As can be seen from Table~\ref{t:tab2}, due to the low statistical quality
of the X--ray data, in many cases the Galactic $n_{\rm H}^{\rm G}$ was 
assumed (in square brackets) in the
spectral fitting. In the cases for which estimates of
$n_{\rm H}$ were derived, they are consistent, within the uncertainties,
with the Galactic values, except for seven GRBs
(GRB970828, GRB980329, GRB980703, GRB990510, GRB000926, GRB001109,
GRB010222) for which the derived $n_{\rm H}$ is significantly higher than
the Galactic value. Since the reported $n_{\rm H}$ values have
been derived assuming only local, Galactic absorption ($z = 0$), 
the true $n_{\rm H}$ may be even higher, 
if it is in the host galaxy or in the ambient medium around the burst.
For example, Weiler at al \cite{Weiler01} (see also the chapter by 
Weiler et al. in \cite{Weiler03}) found that significant amount of 
optical extinction can be ascribed to the burst host galaxy or 
its local surroundings.

\begin{figure}[t]
\vspace{0.5cm}
\centerline{\includegraphics[width=0.7\textwidth]{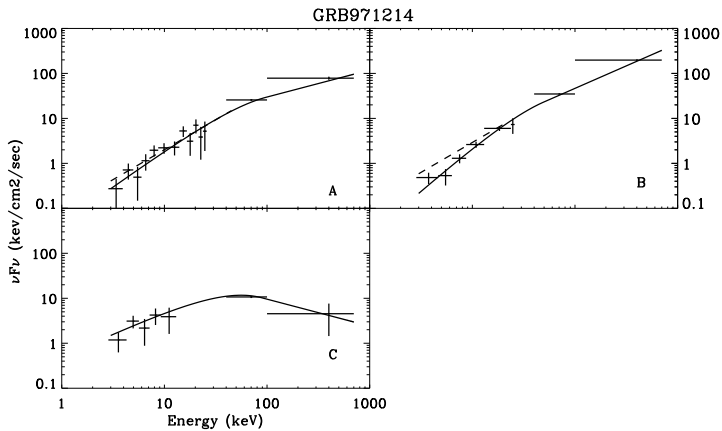}}
\vspace{-6cm}
\caption{Spectral evolution of GRB971214 in three contiguous time intervals, A 
(8~s), B (12~s) and C (18~s), in which the GRB light curve was
subdivided \cite{Frontera00b}.}
\label{f:grb971214_sp}
\end{figure}

The best fit photon indices $I$ range from $\sim - 1.5$ to $\sim - 2.8$ with
a roughly Gaussian distribution with standard deviation
$\sigma_I \approx 0.4$ and mean value $<I > = - 1.95\pm 0.03$,
which is very close to the power law index of the well known synchrotron
source, the Crab Nebula ($I = - 2.1$). The only evidence of spectral 
variability is for GRB970508 and GRB00026 listed in Table~\ref{t:tab2}.
For comparison, the prompt 2--700 keV emission from GRBs with detected
X--ray afterglow is generally well described \cite{Frontera00b} by a
photoeletrically absorbed, smoothly broken power--law \cite{Band93} with
low--energy index $\alpha_1$ (below the break), high energy index
$\alpha_2$ above the break, and peak energy $E_p$ of the $E F(E)$ spectra, all 
which evolve rapidly with time. One relevant feature  of these spectra is
that, at the GRB onset, $E_p$ is generally above the
energy passband of the \sax\ GRBM \cite{Frontera00b} (see the example in 
Fig.~\ref{f:grb971214_sp}).
%
%

Several hints, such as the non thermal shape of the spectra and 
detection of polarized emission from the optical counterparts of
two afterglows (GRB990510 \cite{Covino99,Wijers99} and GRB990712
\cite{Rol00}) point
to synchrotron radiation as the basic emission mechanism 
of the electromagnetic radiation
from GRB afterglows. Also, in the case of the prompt emission,
synchrotron radiation appears to play an important role, even
if other emission processes such as Inverse Compton (IC) may contribute
 at the earliest times (see, e.g., \cite{Frontera00b}). 
Synchrotron radiation is expected within the framework of the fireball
model (see, e.g., reviews by Piran \cite{Piran99,Piran00} and references
therein) as a result of the interaction between a relativistically
expanding fireball and the external medium (''external shock'').
Detailed calculations of the expected synchrotron
spectral shape and its time evolution have been carried out for various
external medium conditions (homogeneous medium \cite{Sari98},
or radially declining density wind \cite{Chevalier99}). 
Also the effect of IC
scattering of low-energy synchrotron photons on shock--accelerated 
electrons has
been investigated (see, e.g., \cite{Panaitescu00,Sari01}).

Following Sari et al. \cite{Sari98}, for the case of a homogeneous medium, 
the instantaneous synchrotron energy spectrum
$F(E)$ is schematically described by a multi--broken power--law, with
four different slopes. At low
frequencies, the flux density $F_\nu \propto \nu^2$ holds up to
the synchrotron self--absorption (SSA) frequency  $\nu_{SSA}$, then 
$F_\nu \propto \nu^{1/3}$ from  $\nu_{SSA}$ 
up to the peak frequency $\nu_m$ ($\nu_m$
corresponds to the minimum--energy of the shock--accelerated electrons).
At still higher frequencies, the spectrum is related to the energy distribution
of the electrons, which is assumed to exhibit a power law shape
$N(E_e) \propto E_e^{-p}$($E_e$ is the electron energy) from  $\nu_m$ up 
to the  cooling
frequency $\nu_c$ ($ \nu_c$ corresponds to the highest energy electrons
that cool more rapidly than the expansion time of the source), by means of 
$F_\nu \propto \nu^{-(p-1)/2}$, while above $\nu_c$,
$F_\nu \propto \nu^{-p/2}$. Depending on the shock evolution,
fully adiabatic or fully radiative, the characteristic
frequencies $\nu_m$ and $\nu_c$ and the maximum flux density $F_m$ vary
in different ways with time. For the fully adiabatic case, 
$\nu_m \propto t^{-1/2}$
and $\nu_c \propto t^{-3/2}$. For the fully radiative case 
$\nu_m \propto t^{-12/7}$ and $\nu_c \propto t^{-2/7}$. 
In the latter case, $F_m$ also changes with
time as $F_m \propto t^{-3/7}$. The different time behaviour of the
characteristic frequencies is thus expected to affect not only the
spectral evolution of the emission but also the time behaviour of the
afterglow light curves (see Section~\ref{s:time}).

The multiwavelength spectrum of the GRB970508 afterglow
fits this scheme very nicely \cite{Galama98a}. Unfortunately, broad band
spectra are available for only a very limited number of GRBs.
An attempt to collect the available data to estimate the 
multiwavelength spectra of GRB afterglows was performed \cite{Masetti99}, 
but the data are still scarce.   Most of the few
available spectra are consistent with the synchrotron shock model
(in addition to GRB970508, GRB971216 \cite{Dalfiume00}, 
GRB980703 \cite{Vreeswijk99a}, and GRB991216 \cite{Frail00}). 
However, in two cases (GRB000926 \cite{Harrison01a} and
GRB010222 \cite{Zand01}), the multiwavelength spectra clearly deviate
from the shape expected from the synchrotron model.  
In both cases, the radio and optical data fit a synchrotron spectrum very
well, but the extrapolation of the spectrum to X-rays is below the measured
X-ray flux density.  Harrison et al. \cite{Harrison01a} interpret this
mismatch as due to the presence of a self--Inverse Compton component.  
The ambient medium density derived is $\sim 30$~cm$^{-3}$, which is higher
than the average InterStellar Medium (ISM) density in a typical galaxy 
($\sim 1$~cm$^{-3}$), but is consistent with a
diffuse interstellar cloud such as those found in star forming regions.

Assuming a synchrotron origin, the photon indices $I$ listed 
in Table~\ref{t:tab2} permit an estimate to be made of the power--law index of
the energy distribution of the electrons accelerated in the
external shocks. Given that most of the  X--ray afterglow
measurements were performed considerably after the main event
(see Table~\ref{t:tab1} for the start times), these can be reasonably assumed
to be taken in the regime of $\nu > \nu_m$, when the peak of the spectrum
had moved toward the optical range. Thus, the electron index $p$ is related to
the photon index by one of  two equations: either $p = - 2 I -1$ or
$p = - 2 I -2$, depending whether the X-ray band is 
between $\nu_m$ and $\nu_c$ (fast cooling) or beyond these frequencies (slow
cooling). 
From the mean value of $I$, we can derive  a mean value 
of the electron index $<p>$ in the range from $\sim 2$ to $\sim 3$,
depending on the cooling regime.

\subsection{Time Behaviour of the X--Ray Afterglow Light Curves}
\label{s:time}
One of the most peculiar properties of the X--ray afterglow light curves
is their power--law decay ($F(t) \propto t^{+\beta}$).
The power--law fading was already noted in the first X-ray afterglow source  
1SAX J0501.7+1146 associated with GRB970228 \cite{Costa97}.
The X--ray fading slope
($\beta_X = -1.33^{+0.11}_{-0.13}$) was consistent with that
($\beta_{opt} =-1.46\pm 0.16$) of the simultaneously discovered optical
counterpart \cite{Jvp97}. 
The power--law decay index and the photon index of the
GRB970228 afterglow 
(see  values in Table~\ref{t:tab2}) were also found to be consistent with
the relationship
predicted by the fireball model and gave strong support
to that model (see e.g., \cite{Wijers97}).
Almost all other X--ray afterglow sources discovered so far show a similar
power--law decay or, in a few cases, a broken
power--law. Table~\ref{t:tab2} reports the best fit temporal indices, $\beta$, 
measured. Apart from GRB980425, which will
be discussed separately below, the power--law indices are distributed
like a Gaussian with mean value
$<\beta> = -1.30\pm 0.02$, and  standard deviation $\sigma_\beta
\approx 0.35$. A relevant feature of the light curves is
that their
extrapolation back to the time of the GRB event is,
in general, consistent with the tail of the prompt X-ray light curve.
This fact is in some cases directly misured (GRB970228 \cite{Costa97} and
GRB970508 \cite{Piro98b}) and in other cases is
inferred from a different approach (see Section~\ref{s:early}).

\begin{figure}[h]
\vspace{-0.3cm}
\centerline{\psfig{figure=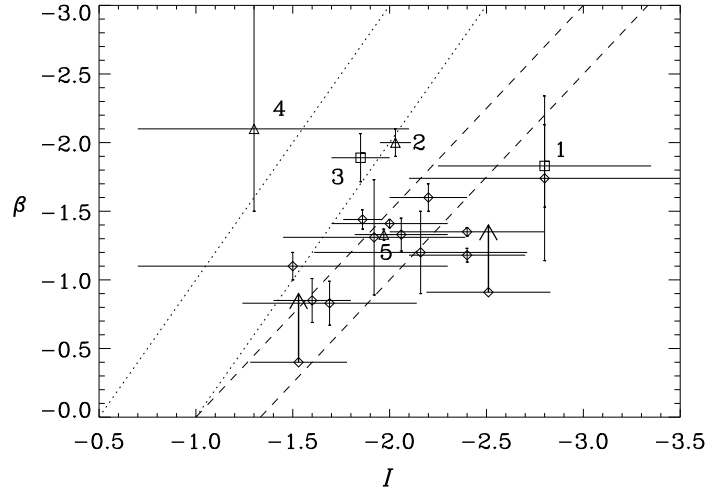,width=10cm}}
\caption{The measured temporal index $\beta$ vs. the photon index $I$ 
of GRB afterglows. The GRBs with evidence of a break in their
X--ray light curves are indicated with triangles, while those with
high temporal index ($\beta <$-1.8) are indicated with squares.
 1: GRB980519, 2: GRB990510, 3: GRB000926, 4: 
GRB010214,  5: GRB010222.
For GRB990510 the temporal index estimated by \cite{Pian01} is quoted.
The region between the dashed lines is for a spherical expansion while the
 region between the dotted lines is for an expanding jet. }
\label{f:beta_vs_i}
\end{figure} 

In a few cases (GRB990510 \cite{Pian00}, GRB010222 \cite{Zand01} and
GRB010214 \cite{Guidorzi03}) comparison of the intensity level of the tail of
the prompt emission with the late afterglow light curve 
(start time several hrs after the GRB event),
shows that a broken power--law may be more consistent with the data. 
In Table~\ref{t:tab2} the power--law indices
and break times of these GRBs are reported. The X--ray breaks  are found to be
consistent with those seen in the optical light curves (see
\cite{Harrison99,Israel99,Stanek99} for GRB990510, and \cite{Masetti01} for
GRB010222).

The interpretation of achromatic breaks in the light curves of GRB
afterglows has been discussed by various authors. In the framework of the
fireball model, a break is expected if the relativistically expanding
material is concentrated within a cone with angular width
$\theta_c$.
As long as the  Lorentz factor $\Gamma$ of the ouflowing material is
larger than $1/\theta_c$, due to  relativistic beaming, the radiation is
emitted within an angle $\theta_b =1/\Gamma$ from the cone axis, with
$\theta_b < \theta_c$.
When $\Gamma$ drops below $1/\theta_c$, the observer begins to see the
edge of the cone and then the effect of the collimated outflow, which causes 
a light curve steepening. 
At the same time ($t_j$), the jet begins to expand sideways,
increasing the rate of decrease of the emitted radiation even more.
This model has been discussed and detailed for various external
conditions by several authors
(see e.g., \cite{Kumar00,Panaitescu99,Panaitescu01,Rhoads99,Sari99a} and
references therein).

Assuming synchrotron radiation, adiabatic expansion and a homogeneous ambient 
medium,
following Sari et al.~\cite{Sari99a}, the energy flux density, if
$\nu_m<\nu<\nu_c$, is given by 
$$F_\nu(t)\propto t^{-3(p-1)/4}$$ 
for $t \le t_j$ and 
$$F_\nu(t)\propto t^{-p}$$ 
for $t \ge t_j$.
However, if $\nu>\nu_c>\nu_m$,
$$F_\nu(t)\propto t^{-3p/4+1/2}$$
 for $t \le t_j$ and 
$$F_\nu(t)\propto t^{-p}$$ 
for $t \ge t_j$. In both cases, after the break the light curve has the
same slope. In this framework, the temporal decay index, $\beta$, 
and the photon index, $I$, are strictly related: 
in the case of spherical expansion
$$\beta = 3I/2 +3/2$$
for $\nu<\nu_c$ and  
$$\beta = 3I/2 +2$$
for $\nu>\nu_c$.
In the case of an expanding jet
$$\beta = 2I +1$$ 
for $\nu<\nu_c$ and  
$$\beta = 2I +2$$ 
for $\nu>\nu_c$.
The above predictions can be easily tested with the available data.

In Fig.~\ref{f:beta_vs_i}, the values of ($I$, $\beta$) for each
afterglow source are shown (see also earlier test \cite{Stratta00}). 
Even though the statistical uncertainties do not allow
us to draw firm conclusions from this diagram, a
clustering of the data  within the region expected from a spherical
expansion is apparent. The above derived mean values of $I$ and $\beta$
confirm this.
Indeed, assuming $I = - 1.95$, in the hypothesis of spherical expansion,
$\beta$ is expected to be in the range from - 0.9 to - 1.4, 
depending on whether
$\nu<\nu_c$ or $\nu>\nu_c$, while in the case of an expanding jet, $\beta$
is in the range from -1.9 to -2.9. This result seems to disagree
with the results of Frail et al.\cite{Frail01} who, by examining
the multiwavelength  light curves of a sample of 17 GRB afterglows find
a break in a large fraction of them which they interpret as evidence
for an expanding jet. They derive the jet opening angle for each of them.
Even though both samples are still too small to infer general properties,
 our result can still be  consistent with those of Frail et al.
\cite{Frail01}. Only 10 of the 17 GRBs appear in Table~\ref{t:tab2} and,
for most those 10, either the derived X--ray parameters refer to epochs before
the breaks reported by Frail et al. or the X--ray light curves are of poor 
statistical quality.

As can be seen from Fig.~\ref{f:beta_vs_i}, the position of three events 
is consistent with the jet region between the dotted lines.
Two show evidence of a break in their X--ray light curves
(GRB010510, GRB010214), and one (GRB000926) exhibits a high temporal index
(but it shows a break in the optical light curve \cite{Piro01a}).
The X--ray light curve of GRB990510  was discussed by Pian et al.\cite{Pian01}, 
who found it consistent with the jet model and derived the following
parameters of the jet and circumburst medium: ambient density of
0.13~cm$^{-3}$, electron energy distribution index of 
$p = 2$, isotropic equivalent 
energy of $E_0 = 5\times 10^{53}$~erg, and opening angle of 
$\theta_c =2.7^\circ$.

In the case of GRB010222, which is within the spherical expansion
region of Fig.~\ref{f:beta_vs_i} but shows a break consistent with 
that observed in the optical band \cite{Masetti01}, in't Zand et al.
\cite{Zand01} interpret it as probably due to the
transition of a spherically expanding fireball from a relativistic to a
non-relativistic phase (NRP) \cite{Dai99}. Also in this case a break is 
expected when
the shock wave has swept up a rest-mass energy equal to its initial energy
\cite{Livio00,Piro01a,Wijers97}. 
Assuming synchrotron radiation from an adiabatically cooling shock as
above,
the change in the temporal index is now  shallower
than for the case of an expanding jet. After the transition to NRP, 
we have 
$$\beta = 3I +18/5$$
for $\nu<\nu_c$ and 
$$\beta = 3I +5$$
 for $\nu>\nu_c$. 
In the case of the GRB010222, in spite of the fact that the measured values
of $\beta$ and $I$ after the break are still inconsistent with each
other assuming NRP, this model  predicts values of the electron index $p$,
when derived from $\beta$  or $I$, which are much more consistent with
each other than in the case of an expanding jet ($-1.94 \pm 0.10$ vs.
$-2.22\pm 0.03$ for transition to NRP against $-1.33\pm 0.03$ vs. 
$-1.94\pm 0.10$ for an expanding jet) \cite{Zand01}.
Assuming the NRP model, from the break time estimate, a high
ambient gas density is inferred ($\sim 10^6$~cm$^{-3}$ \cite{Zand01}), which
could be consistent with the fact that a Comptonized component is
needed by the X-ray afterglow spectral data (see Section~\ref{s:spectrum}).

The existence of high temporal indices ($\beta \le -1.8$, 
see Table~\ref{t:tab2}) has also been taken  as evidence for beamed
emission (GRB980519 \cite{Sari99a}) or for a transition to NRP
(GRB000926 \cite{Piro01a}). However, in the case of GRB980519, the X--ray
data are  consistent with a spherical expansion (see
Fig.~\ref{f:beta_vs_i}), and in the case of the GRB000926 afterglow, the
observational scenario appears more complex. By combining together optical and
X--ray observations, two
breaks are apparent \cite{Piro01a}. Piro et al.\cite{Piro01a} interpret
the early break as due to a relativistic transition from a spherical to a 
jet--like phase, and the later break as due to the transition 
of the jet to NRP \cite{Livio00}.
This interpretation implies the presence of a
moderately collimated outflow ($\theta_c \approx 27^\circ$) that expands
in a dense medium ($n \approx 4 \times 10^4$~cm$^{-3}$). Unfortunately 
the derived medium
density  is much higher than that ($n\sim 30$~cm$^{-3}$)
inferred from the broad band spectrum \cite{Harrison01a},  as discussed in
Section~\ref{s:spectrum}. Also, in the case of GRB010222 a late break
could  be the explanation of the inconsistency of the 
\sax\ light curve with the \emph{CHANDRA} flux point taken 9 days after the
GRB \cite{Zand01}.


\begin{figure}[t!]
\centerline{\psfig{figure=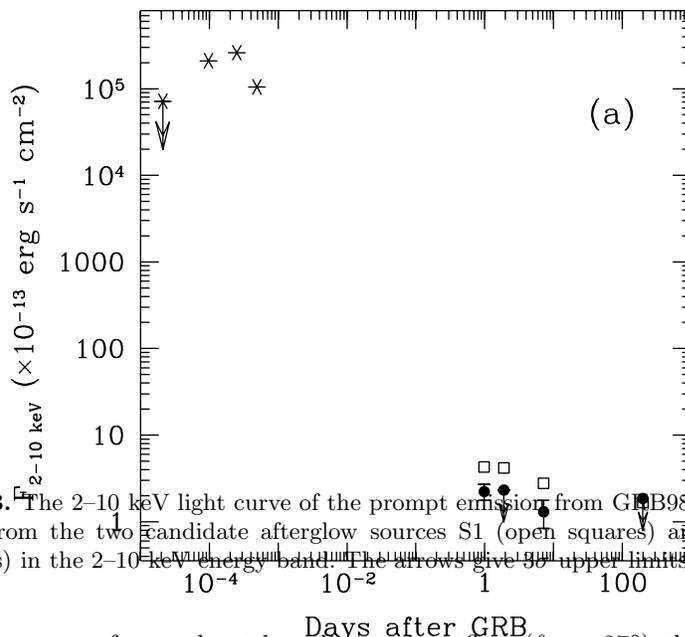,width=10cm}}
\vspace{-3cm}
\caption{The 2--10 keV light curve of the prompt emission from GRB980425 
(stars) and from the two candidate afterglow sources S1 
(open squares) and S2 (filled circles)
in the 2--10 keV energy band. The arrows give 3$\sigma$ 
upper limits \cite{Pian00}.}
\label{f:grb980425_lc}
\end{figure} 

The presence of beamed emission would help to reduce the dramatic GRB
energy problem. While the isotropic gamma--ray energy
$E_{iso}(\gamma)$ ranges from $5 \times 10^{51}$ to 
$3 \times 10^{54}$~ergs (see e.g.,
\cite{Amati02a,Bloom01,Schmidt99}),
the $E_\gamma$ for a jet with opening angle
$\theta_c$ would be a factor of 2/$\theta_c^2$ lower. Frail et al. 
\cite{Frail01} attribute the range of
variation of $E_{iso}(\gamma)$ to a range of jet cone angles and give an
estimate of the mean value of $E_\gamma \sim 5\times 10^{50}$~ergs, where a
conversion efficiency of the jet kinetic energy into electromagnetic
radiation of 0.2 is assumed \cite{Guetta01}. Rossi et al. \cite{Rossi02}
assume a jet as well, but with a variation in the output energy which
depends on the offset angle from the jet axis ($\propto \theta^{-2}$) and
attribute the dispersion in the $E_{iso}(\gamma)$ energy to differences in 
the jet axis-to-observer viewing angles. 

The afterglow of GRB980425 deserves a separate discussion, because of
its large impact on the GRB--supernova connection issue. The GRB
occurred at a time and in a position  which are both  consistent with those
of the peculiar Type Ic supernova SN1998bw in the nearby galaxy
ESO 184-G82 ($z=0.0085$) \cite{Galama98b,Iwamoto98,Kulkarni98}. Following
the detection with \sax\ \emph{WFCs} and \emph{GRBM} (see Table~\ref{t:tab1}), 
a prompt X--ray observation of the GRB error box led to the discovery of two
weak sources, S1 and S2, within the \emph{WFC} error box, one of which,
S1, was consistent with the SN1998bw position \cite{Pian00}. The 2--10 keV flux
level of S2, about two times lower than that of S1, was at the sensitivity 
limit of the
\sax\ \emph{MECS} telescopes so that even very small variations of this source 
could lead to a positive detection or an upper limit. Indeed, the source was no
longer detected one day after the first observation and it was barely 
detectable agains 6 days later. 
Finally, it was not detected during an observation in November 1998 
(see Fig.~\ref{f:grb980425_lc}).
The source S1 was stronger, but showed a very weak power--law
fading with index $\beta = - 0.2$. Neither S1 nor S2 showed ''typical'' 
fading behaviour, illustrating the peculiar nature of the burst and afterglow.
Further observations of the GRB980425 field with more
sensitive X--ray instrumentation are of key importance. An observation 
performed with XMM--Newton \cite{Pian02} has shown that the X--ray flux from S1
has further faded from the last \sax observation according 
to an exponential law, while S2, still visible, is the superposition of more 
sources, likely AGNs.

\subsection{Emission/Absorption Features in the X--Ray Spectra}
\label{s:features}
\begin{figure}{t}
\centerline{\psfig{figure=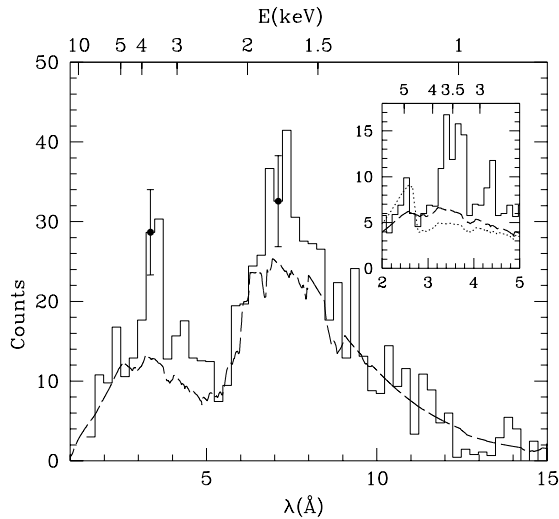,width=8cm}}
\caption{Spectrum of the X-ray afterglow of GRB991216 obtained
with the {\it CHANDRA} high energy gratings. The 4.7$\sigma$ excess
at 3.5 keV is apparent. In the inset the fine structure of the line is shown
 \cite{Piro00}.}
\label{f:grb991216_sp} 
\end{figure}

Evidence of emission features has been reported in the X-ray 
afterglow spectra of GRB970508 \cite{Piro99a}, GRB970828
\cite{Yoshida99}, GRB991216 \cite{Piro00}, and GRB000214
\cite{Antonelli00} \footnote{After this paper was written, 
evidence of 5 emission features from GRB011211 (Mg XI, Si XIV, S XVI, Ar XVIII, Ca XX)
were reported \cite{Reeves02}, but their reality is questioned 
\cite{Borozdin03,Rutledge03}.}. Table~\ref{t:tab3} summarizes the major properties of
these features and Fig.~\ref{f:grb991216_sp} shows one of the two features
discovered in the GRB991216 X-ray afterglow.
%
%
A feature at 3.5 keV (97\% confidence level)
was observed from GRB970508 only during the first part (10 hrs) of the
follow--up observation of this burst, before
a delayed outburst occurred in both the X-ray and optical bands \cite{Piro98b}.
Also in the case of GRB970828 an emission feature at 5 keV
(confidence level of 98.3\%) was observed for a
limited time duration ($\sim 5.5$~hrs) during a
flare activity of the GRB afterglow. The two emission features from the
GRB991216 afterglow (4.7$\sigma$ excess above the continuum emission
at 3.5 keV, 2$\sigma$ excess for the feature at 4.4 keV) were
detected for the entire observation duration (3.4~hrs) performed 37~hrs
after the main event with \emph{CHANDRA}. 
In the last case (GRB000214), the
continuum level was observed to fade while the emission feature at 4.7 keV
(3$\sigma$ significance level) remained stable for the entire duration
of the observation ($\sim 29$~hrs).
Taking into account that three of these GRBs have  host
galaxies with known redshifts (0.835 for GRB970508 \cite{Metzger97},
0.9578 for GRB970828 \cite{Djorgovski01b} and 1.02 for GRB991216
\cite{Vreeswijk99b}), the line features in their respective
co-moving frames are consistent with Fe fluorescence lines
(Fe {\sc i}--Fe {\sc xx} K$\alpha$ line in the case of GRB970508, 
Fe K line of H--like ions in the case of the lower energy line from
GRB991216) or Fe recombination edges (recombination of Fe H--like ions
for both the line feature from GRB970828 and the higher energy
feature from GRB991216). In the case of GRB000214, for which the redshift is
not known, if the feature is interpreted as an Fe K fluorescence line,
the implied redshift of the GRB source ranges from 0.37 to 0.47,
depending on the ionization status of Iron, from neutral to hydrogen--like
Fe, respectively.

Independent of the specific identification, all the detected lines
point to the presence of ionized Fe at the time of the afterglow
measurements, with the ionizing radiation likely due to the GRB flash.
The line widths (see Table~\ref{t:tab3}) are not accurately determined
except in the case of the 3.5~keV line from GRB991216  (see
Fig.~\ref{f:grb991216_sp}). In this case the
line width has been interpreted \cite{Piro00} as being of kinematic origin,
implying velocities of the line emitting material of $\sim 0.1c$. This
velocity would also be that of the material ejected by the GRB progenitor (if
a supernova) a few months  before the main GRB event.
The large equivalent widths of the lines favour a reflection geometry 
for the line emitting material \cite{Vietri01}.
Indeed, for a reflection process the ionizing flux is always efficiently 
reprocessed into line photons by the outermost Fe layer (optical depth
$\tau_{Fe}\sim 1$), while for a transmission process an efficient
re-processing of the ionizing flux would require a particular tuning of the
density of both the scattering material and free electrons. The amount of
Fe needed mainly depends on the density of the scattering
cloud and on its distance from the GRB site. In the reflection model worked
out by Vietri et al. \cite{Vietri01}, the required iron mass to obtain the
observed line fluxes is significant (a sizable fraction of a solar mass in
the case of GRB991216) which is not expected even for the largest known stars.
Only a supernova remnant  could contain so much iron.
However in the model by Rees and M\'esz\'aros \cite{Rees00}, in which 
a continuing (hrs to days) power output after the burst 
in a magneto-hydrodynamic (MHD) wind is
required, it is the interaction of this wind with the walls of a funnel
previously created by the GRB itself (assumed to be produced in a
collapsing star) that gives rise to the
emission features. They result from a reprocessing of part of the wind
luminosity by a thin layer of the funnel. In this case, the  iron
mass needed to obtain the line flux is much lower ($\sim 10^{-8}M_\odot$ for
GRB991216), thanks to the high material density and the high
recombination rate of the Fe ions. In this model the GRB environment
is not required to be enriched in Fe by a previous supernova explosion.
In a later paper, M\`esz\`aros and Rees \cite{Meszaros01b} proposed 
a different model for the emission line in which the total amount of
Fe is significantly larger ($\sim 10^{-5}$~M$_\odot$). Here, a hypernova
progenitor releases a relativistic
shock when the inner black hole is sufficiently powerful to drive the
shock progress into the outlying star layers. The heated matter 
(heavily contaminated by the Fe nuclear matter), which is 
not accelerated to relativistic energies, expands outside the main body 
of the star in a cocoon and produces the Fe line thermally. 

In this context, the discovery of a transient absorption feature (13~s
duration) at 3.8 keV in the
prompt emission of GRB990705 \cite{Amati00} could provide an important
datum for  settling the issue of the Fe abundance and its mass and
location within the environment in which the emission lines are produced.
Assuming that the observed absorption feature was due to a redshifted Fe
K--edge \cite{Amati00} (consistent with the optical redshift $z_{opt} =
0.8424 \pm 0.0002$ of the GRB host galaxy \cite{LeFloc02}), 
the derived Fe abundance
($A_{Fe} \sim 75$) is typical of a supernova remnant (SNR) and its transient
nature could be a consequence of
its photoionization by the GRB photons. From the estimated column density,
supposing a uniform distribution of SNR material, its
distance from the GRB site was estimated to be $\sim 0.1$~pc.
Such a short distance makes it very likely that the GRB site
was coincident with that of a supernova explosion. The time for the remnant
to grow to this size was estimated as $\sim$ 10 ~yrs.
A drawback of this scenario \cite{Boettcher01,Lazzati01a} is the huge
amount of Iron (tens of solar masses) required, unless the absorbing iron
is clumped and a clump is by chance along the line of sight.

Lazzati et al. \cite{Lazzati01a} propose an alternative model that does
not require a large Fe mass. In this model the assumption is
that the quasi--complete ionization of iron is achieved in a much 
shorter time than the duration of the feature (13~s), which is due
to resonant scattering of the GRB photons off H-like iron
(transition 1s-2p, $E_{\rm rest} = 6.927$~keV), while its disappearance
is due to a halt of the electron recombination by an electron temperature
increase. In this model, to fit the
data, a Fe relative abundance $\sim 10$ times the solar value and an Fe
mass of $\sim 0.2 M_{\rm \odot}$ are required.
Also in this model, an iron rich environment, typical of an SNR, is required
and the distance to the absorbing iron is low
($\sim 2 \times 10^{16}$~cm). As a consequence, independent of the specific
model, a SN explosion preceding the GRB event is required, which is likely
to be associated with the GRB event itself.
In this framework, the afterglow emission features appear when the
strong ionizing flux has decreased and electron recombination restarts.

\section{The Early X--Ray Afterglow}
\label{s:early}
A peculiar property of the X-ray afterglow light curves is their very
smoothly fading behaviour, except in a few cases in which an outburst 
(GRB 970508), a flaring activity (GRB 970828), and a rise from below the \sax\ 
\emph{MECS} sensitivity (GRB 981226), were observed
\cite{Frontera00a,Piro98b,Yoshida99}.
By comparison, the time profile of the prompt $\gamma$-ray emission is erratic
and even spiky. 
In order to explain the complex time profiles of GRBs, it has been hypothesized
\cite{Kobayashi97,Sari97a,Sari97b} that the GRB itself could be due to 
internal shocks, with fast-rise exponential-decay (FRED) events perhaps also
involving external shocks \cite{Dermer99}). However, there is a general
consensus that the late afterglow emission is likely due to external shocks
(see, e.g. \cite{Meszaros97,Sari98}). In this scenario, 
the two phenomena can evolve in different ways and can involve
energetics which are not necessarily correlated. It has also been suggested
\cite{Sari97c,Sari99b} that the early afterglow could start already  only a few
tens of seconds after the burst. In fact, in a few cases
like GRB970228 \cite{Costa97,Costa00b} and  GRB970508 \cite{Piro98b},
the extrapolation of the 2-10~keV afterglow fading law back to the
time of the GRB main event gives a flux consistent with that measured
during the last portion of the event. Using the ratio between the 2-10 keV
fluence of the last portion of the prompt emission and that of the late
afterglow, Frontera et al. \cite{Frontera00b}  demonstrated that this property
is valid also for an entire sample of GRBs with known afterglows; approximately
the second half of the GRB prompt X--ray emission is strictly correlated
with the afterglow emission. This suggests that at least the tail of
the GRBs can be
due to afterglow emission, while at the  onset of the primary event this
contribution is negligible. Once the rise of the afterglow has been
established, the framework of the fireball model permits the initial 
Lorentz factor
$\Gamma_0$ of the shocked ambient medium to be evaluated 
\cite{Sari99b}. From the above results $\Gamma_0 \sim 150$
for all GRB events investigated, except for GRB980425 in which  a much lower
value ($\Gamma_0<50$) is found \cite{Frontera00b}.

At higher energies ($>$ 10~keV), as mentioned in Section~\ref{s:spectrum}, 
late afterglow emission has been discovered only from GRB990123 with the \sax\ 
\emph{PDS} instrument \cite{Heise02}. However, early  afterglow
emission is also expected from, e.g., electron--synchrotron 
self-Inverse Compton 
and proton--synchrotron processes \cite{Sari01,Zhang01} and some evidence 
of a possible early, high energy afterglow
has been reported \cite{Burenin99,Connaughton01,Giblin99}.
The most suggestive result is that found by Giblin et al. \cite{Giblin99}
with the strong \emph{BATSE} event GRB980923 in the 25--300~keV range. 
The \emph{BATSE} data show a GRB light curve which,
after 40~s of erratically variable flux, smoothly decays with a power--law
shape of index $\beta \sim -1.8$ for about 400~s 
until the event becomes invisible. 
Correspondingly, the energy spectrum shows a power--law shape
with a photon index $I$, which, after 40~s from the GRB onset,
exhibits a jump from  -1 to -1.7, which is typical 
of the X--ray afterglow photon indices (see Section~\ref{s:spectrum}).

\section{Open Issues and Future Prospects}
\label{s:issues}

In spite of the tremendous advances in the years since the discovery 
of the first X-ray afterglows, many questions about
the GRB phenomenon are still open which can only be answered with
further X--ray observations.
Analysis of the afterglow radiation seems to indicate a synchrotron 
origin, but in some cases multiwavelength spectra
(see Section~\ref{s:spectrum}) show the presence of an
additional component in the X--ray band interpreted as electron
self--Inverse Compton radiation.
It is important to obtain good quality, multiwavelength spectra for a large
sample of GRBs to determine the relevant emission mechanisms of the afterglow 
radiation.

Breaks in the X--ray afterglow light curves have been observed for a few GRB
afterglows, while at longer wavelengths such breaks are more frequent
\cite{Frail01}. The jet model predicts that these breaks are achromatic if
the jet propagates into a medium with uniform density. However, if the jet
moves into a wind-like circumstellar environment ($\rho \propto r^{-s}$ 
with $s\approx 2$), the break is
expected to be less relevant in X-rays than in the optical band
\cite{Kumar00}. In addition, other models have been proposed for interpreting
such breaks (e.g., transition to NRP, see Section~\ref{s:time}). Thus it is
of key importance to measure X-ray afterglow light curves with high
statistical quality in order to test the
jet model and to derive the ambient medium properties to 
be compared with the properties derived from the multiwavelength spectra.

The ambient medium
density, along with other GRB environmental properties (composition, velocity,
temperature, metal abundance) can also be inferred from the  detection of
X--ray lines or edges in the spectra of the prompt and delayed X-ray emission.
This information is important for determining the nature of the GRB progenitor,
which is one of the most debated issues in astrophysics today. 
For the case of collapse of a
massive progenitor, the GRB environment is expected to be Fe polluted and
line features are more likely.
On the contrary in a clean environment such as that
expected for neutron star mergers, Fe lines are not likely.
The X--ray lines detected so far point toward the presence of a
medium highly ionized by the GRB main event, with iron abundance 
higher than solar. It is likely that this iron is produced and 
ejected by the GRB progenitor before the burst, which supports the hypothesis 
a supernova-like explosion
(hypernova, collapsar or supranova 
\cite{Paczynski98,Vietri98,Woosley93}). However many questions are still
open even in this scenario. For example:
\begin{itemize}
\item[-] Why do we not see any absorption feature
in the prompt emission of the GRBs which show an
emission feature during the afterglow phase (GRB970508 or GRB000214)?
\item[-] Why are the emission lines in the afterglow spectra 
of GRB970508 and GRB970828
observed only for a limited amount of time, while in the case of GRB000214
the line is stable? 
\item[-] Why in many other GRBs observed with \sax\ do we not see any feature? 
\item[-] Is this due to the limited sensitivity of
the \sax\ instrumentation or is it due to an intrinsic physical process?
\end{itemize}
 
Only more sensitive and continuous observations of GRBs from the main event 
through to the late afterglow can answer these questions.

High quality light curves of the GRB prompt emission and afterglows, with
no interruption, could also resolve the issue of the  onset time of
the afterglow emission phase, its time profile and its spectral evolution.
This
information is of key importance to study the feedback of the intense
$\gamma$--ray flux on the dynamics of the fireball  and on the afterglow
emission properties \cite{Beloborodov02,Meszaros01a,Thompson00}. High
quality light curves at multiple wavelengths could definitely settle the
question of the rate of GRBs with single or multiple broken power--law temporal
behaviour. If the breaks in the light curve imply the presence of jets, 
the GRB energetic problem could be solved, or mitigated, and
ordinary supernovae could be energetic enough to be the origin 
of GRB events \cite{Frail01}.
High quality light curves would also permit the study of short time
variations in the afterglow fading, and thus of the degree of 
homogeneity of the circumburst medium \cite{Frontera00a}.

Another open issue concerns the origin of short ($<$1~s) GRB events.
Until now, no prompt (within few hrs) follow-up of these events has been
possible and no afterglow source has ever been identified \cite{Hurley02}.
A claim for the detection of a transient (100~s duration) and fading hard
X--ray emission,  obtained by summing the \emph{BATSE} light curves of a sample
of 76 short GRBs has recently been reported (see \cite{Lazzati01b}, but also 
\cite{Connaughton01}). Lazzati et al. \cite{Lazzati01b} interpret this
emission as possible evidence for afterglows from short GRBs. This claim 
still has to be confirmed by other, more sensitive observations and its
origin is not clear. It could also be prompt emission from truly long
GRBs with have a spectrally hard, short and strong pulse at their onset, and a
long tail which is not individually observable because it is below
the sensitivity limit of the instrumentation used.
More sensitive GRB observations with immediate position identification 
are needed to disentangle the short GRB origin issue and the
afterglow properties of these events.

The cause of ''dark'' GRBs (see also the chapter by E. Pian) is still a mystery.
A possible explanation is the presence of
dust in the environment around GRB sites that obscures the optical
emission, even if the GRB output may be enough to destroy the dust in
the immediate surroundings \cite{Djorgovski01a}. Observations of X--ray
lines from these GRBs could be crucial for estabilishing the
environmental  properties. However other possibilities cannot be
excluded, such as a rapid decline of the optical light or a high redshift of
the object. Only fast afterglow searches, starting minutes
after the main event with multiple wavelength coverage of the afterglow
emission, can definitely settle this issue.

The X--ray flashes
detected with \sax\  \cite{Heise01} with no $\gamma$-ray counterparts
are still a mystery. They could
be GRBs at very high redshifts (z$>$10) or GRBs with very soft spectra, or GRBs
from fireballs with a low Lorentz factor. More sensitive observations of these
events are needed.

The fireball model of GRBs is currently the most successful for
interpreting the X--ray data. Other models have been proposed, such as 
the cannonball (see, e.g.,\cite{Dar00}) and only future
observations will be able to distinguish between, and determine
the validity of, these alternative models.

One of the major limitations of the current investigations on GRB afterglows
is the delay from the main event to the follow-up observations. Many of the
issues discussed herein will be solved when future missions, now under
development (in particular {\it Swift Gamma-Ray Burst Explorer} ({\it Swift})
 \cite{Gehrels00} and {\it Gamma-Ray Large Area Space Telescope}({\it GLAST}) 
\cite{Gehrels99}) are operating.
{\it Swift} will perform immediate ($\sim$ 1 min) follow-up
observations in the X--ray and optical bands and will trigger very
prompt optical, near-infrared, and radio observations with ground-based
 telescopes.  The $\gamma$--ray mission {\it GLAST} will allow a
sensitive and high time resolution study of the highest energies (GeV)
emitted by GRBs and other missions, like the European 
{\it International Gamma-Ray Astrophysics Laboratory} ({\it INTEGRAL}) 
(se, e.g., \cite{Mereghetti01}) 
and the Italian {\it Astro-rivelatore Gamma a Immagini LEggero} 
({\it AGILE}) \cite{Tavani01} are expected to make key 
contributions to the resolution of the GRB mystery.

\section{Acknowledgments}
I wish to thank Lorenzo Amati for his generous contribution to the preparation
of the Tables 1 and 2 of this paper, Mario Vietri for his very
useful comments and suggestions for improving the paper, and Nicola Masetti and
John Stephen for their careful reading of the manuscript. 
I also wish to thank
the colleagues of the \sax\  team (L. Amati, C. Guidorzi, L. Nicastro,
L. Piro, P. Soffitta), who have permitted the publication of 
preliminary results.
Many thanks expecially to my wife for her sympathy for my days 
of work at home for this chapter,
when I should have been available for family needs.
I also like to acknowledge financial support from the 
{\it Italian Space Agency} {\it (ASI)} and from the Ministry 
of the University of Italy (COFIN funds).

\begin{table}[h!]
\caption{GRBs Observed for X--rays Afterglows}
\renewcommand{\arraystretch}{1.3}
\setlength\tabcolsep{1.5pt}
\scriptsize
\begin{tabular}{lllllllll}
\hline\noalign{\smallskip}
  GRB & Instr. & Error & Start of & Instr. & X--ray  & Source  & Error &  Ref. \\
 & for & box & X--ray & for  & counterpart & name$^{\mathrm e}$ &  box & \\
 & GRB & radius & follow- & follow- & flux at start$^{\mathrm d}$ & & radius$^{\mathrm e}$ & \\
 & loc.$^{\mathrm a}$ & & up$^{\mathrm b}$& up$^{\mathrm c}$ & (erg~cm$^{-2}$~s$^{-1}$) & & &  \\
\noalign{\smallskip}
\hline
\noalign{\smallskip}

 970111 & WFC & 3' & 16.5 & NFI & $< 1.6\times 10^{-13}$ & 1SAXJ1528.1+1937(?) &  & \cite{Feroci98a} \\
 970228 & WFC & 3' & 8.27 & NFI & $3.2\times10^{-12}$ & 1SAXJ0501.7+1146 & 50'' & \cite{Costa97} \\
        &     &    & .... & HRI & $\sim2 \times10^{-12}$ & RXJ050146+1149.9 & 10'' & \cite{Frontera98a} \\
 970402 & WFC & 3' & 8.39 & NFI & $2.7\times10^{-13}$ & 1SAXJ1450.1-6920  & 50'' & \cite{Nicastro98a} \\
 970508 & WFC & 1.9' & 5.6 & NFI & $\sim 7 \times10^{-13}$ & 1SAXJ0653.8+7916 & 50'' & \cite{Piro98b,Piro99a} \\
 970616$^{\mathrm f}$ & BAT & 2$^\circ$ & 4.0 & PCA & $ 1.1\times10^{-11}$ & 01 18 57 -05 28 00  & 0.7' $\times$18'& \cite{Marshall97a} \\
 970815 & ASM & 6'$\times$3' & 76.8 & HRI & $< 1.0\times 10^{-13}$ &  &  & \cite{Greiner97} \\
 970828 & ASM & 2.5'$\times$1' & 3.6 & PCA & $1.15\times10^{-11}$ & 18 08 32.2 +79 16 02 & 30'' & 
\cite{Marshall97b,Murakami97a,Murakami97b,Yoshida99} \\
 971214 & WFC & 3.3' & 6.55 & NFI & $\sim 7 \times 10^{-13}$ & 1SAXJ1156.4+6513  & 60'' & \cite{Dalfiume00} \\
 971227 & WFC & 8' & 14.4 & NFI & $2.6\times10^{-13}$ & 1SAXJ1257.3+5924 & 90'' & \cite{Antonelli99a} \\
 980329 & WFC & 3' & 7.0 & NFI & $1.4\times10^{-12}$ & 1SAXJ0702.6+3850  & 50'' & \cite{Zand98} \\
 980425 & WFC & 8' & 10.2 & NFI & $4.0\times10^{-13}$ & 1SAXJ1935.0-5248  & 90'' & \cite{Pian00} \\
 980515 & WFC & 4' & 9.83  & NFI & $1.6\times10^{-13}$ & 1SAX J2116.8-6712  & 90'' & \cite{Feroci98b} \\
 980519 & WFC & 3' & 4.74 & NFI & $6.0\times10^{-13}$ & 1SAXJ2322.3+7716  & 50'' & \cite{Nicastro98b,Nicastro99a} \\ 
 980613 & WFC & 4' & 8.63 & NFI & $3.3\times10^{-13}$ & 1SAXJ1017.9+7127 & 50'' & \cite{Costa98,Soffitta01} \\
 980703 & ASM & 4' & 22.0 & NFI & $4.7\times10^{-13}$ & 1SAXJ2359.1+0835 & 50'' & \cite{Vreeswijk99a} \\
 980706$^{\mathrm g}$ & COM & 4$^\circ$ & 2.7 & PCA & $4.0\times10^{-11}$&  &  & \cite{Marshall98,VanParadijs00} \\
 981226$^{\mathrm h}$ & WFC & 6' & 11.3 & NFI & $< 1.5\times 10^{-13}$ & 1SAXJ2329.6-2356 & 60'' & \cite{Frontera00a} \\
 990123 & WFC & 2' & 5.8 & NFI & $1.6\times10^{-11}$ & 1SAXJ1525.5+4446 & 50'' & \cite{Heise99,Heise02} \\
 990217 & WFC & 3' & 6.5 & NFI & $< 1.0\times 10^{-13}$ & & & \cite{Piro99b} \\
 990506 & BAT & 9.6' & 3.0 & PCA & $3.2\times10^{-11}$ & 11 54 46 -26 45 35 & 6' & \cite{Hurley99} \\  
 990510 & WFC & 3' & 8.0 & NFI & $5.2\times10^{-12}$ & 1SAXJ1338.1-8030 & 56'' & \cite{Kuulkers00a} \\
 990627 & WFC & 3' & 8.0 & NFI & $3.5\times10^{-13}$ & 1SAXJ0148.5-7704 & 69'' & \cite{Nicastro99b} \\
 990704 & WFC & 7' & 8.3 & NFI & $8.0\times10^{-13}$ & 1SAXJ1219.5-0350 & 60'' & \cite{Feroci01} \\
 990705 & WFC & 3' & 11.38 & NFI & $1.9\times10^{-13}$ & 1SAXJ0509.9-7207 & 120'' & \cite{Amati00} \\
 990806 & WFC & 3' & 7.77 & NFI & $4.0\times10^{-13}$ & 1SAXJ0310.6-6806 & 60'' & \cite{Frontera99,Montanari01} \\
 990907$^{\mathrm i}$ & WFC & 6' & 10.98 & NFI & $1-2\times10^{-12}$ & 1SAXJ0731.2-6928 & 180'' & \cite{Piro99c} \\
 991014 & WFC & 6' & 11.76 & NFI & $4.0\times10^{-13}$ &  1SAXJ0651.0+1136 & 90'' & \cite{Zand00} \\
 991106$^{\mathrm j}$ & WFC & 3' & 7.8 & NFI & $1.25\times10^{-13}$ ? & 1SAXJ2224.7+5423 ? & 90'' & \cite{Antonelli99b} \\
 991216 & BAT & 1.1$^\circ$ & 4.03 & PCA & $1.24\times10^{-10}$ & 05 09 31.5 +11 17 05.7 & 1.5'' & \cite{Piro00,Takeshima99c} \\
 000115 & BAT & 1.7$^\circ$ & 2.92 & PCA & $4.3\times10^{-11}$ & 08 03 12 -17 05 59 & 18' & \cite{Marshall00b} \\ 
 000210 & WFC & 2' & 7.2 & NFI & $4.5\times10^{-13}$ & 01 59 15.7 -40 39 32.3 & 1.6'' & \cite{Costa00,Germire00,Piro02} \\ 
 000214 & WFC & 6.5' & 12.0 & NFI & $7.7\times10^{-13}$ & 1SAXJ1854.4-6627 & 50'' & \cite{Antonelli00} \\
%
%
 000528$^{\mathrm k}$ & WFC & 2' & 8.3 & NFI & $1.8\times10^{-13}$ & 1SAXJ1045.1-3359 & 90''& \cite{Kuulkers00b} \\
 000529$^{\mathrm l}$ & WFC & 4' & 7.5 & NFI & $2.8\times10^{-13}$ & 1SAXJ0009.8-6128 & 120'' & \cite{Feroci00} \\
 000615$^{\mathrm m}$ & WFC & 2' & 10.04 & NFI & $6.7\times10^{-14}$ & 1SAXJ1532.4+7349 & 120'' & \cite{Nicastro01} \\
 000926 & IPN & 3'$\times$10' & 48.96 & NFI & $3.8\times10^{-13}$ & 17 04 01.6 +51 47 08.6 & 2'' &
 \cite{Harrison01a,Piro01a} \\
 001025$^{\mathrm n}$ & IPN & 5' & 45.60 & XMM & $1-2\times10^{-13} $ & 08 36 35.42-13 04 09.9 & $<$10''  & \cite{Altieri00a,Altieri00b} \\
 001109 & WFC & 2.5' & 16.5 & NFI & $7.1\times10^{-13}$ & 1SAXJ1830.1+5517 & 50'' & \cite{Amati02b} \\
 010214 & WFC & 3' & 6.28 & NFI & $1.0\times10^{-12}$ & 1SAXJ1741.0+4834 & 60''&\cite{Guidorzi03} \\
 010220 & WFC & 4' & 16.0 & NFI & $<1.0 \times10^{-13}$ & & & \cite{Manzo01} \\
 010222 & WFC & 2.5' & 8.0 & NFI & $1.2\times10^{-11}$ & 1SAXJ1452.2+4301 & 30''& \cite{Zhang01}\\
 011121 & WFC & 2' & 21.25 & NFI & $\sim 1.0\times10^{-12}$ & 1SAXJ113426-7601.4 & 50'' & \cite{Piro01b,Piro03}\\
 011211$^{\mathrm o}$ & WFC & 1' & 11.13 & XMM & $1.9\times10^{-13}$ & 11 15 17.9-21 56 57.5 &$ <$10''& \cite{Reeves02} \\
\hline
\end{tabular}
\end{table}

\setcounter{table}{0}
\begin{table}[h!]
\caption{(continued)}
\renewcommand{\arraystretch}{1.3}
\setlength\tabcolsep{1.5pt}
\scriptsize

$^{\mathrm a}$ \emph{WFC} on board \sax; \emph{BAT} = \emph{BATSE} on board \emph{CGRO}; \emph{ASM} on board \emph{RXTE};
\emph{COM} = \emph{COMTEL} on board \emph{CGRO}; 
\emph{IPN} = InterPlanetary Network. \\
$^{\mathrm b}$ Earliest afterglow search, in hours from the GRB detection. \\
$^{\mathrm c}$ Instrument used for the earliest afterglow search: \emph{NFI}
 on board \sax; \emph{HRI} on board \emph{ROSAT}; \emph{PCA} on 
board \emph{RXTE}; \emph{XMM} = \emph{XMM-Newton} satellite. \\
$^{\mathrm d}$ Energy band: 2-10 keV. All upper limits are 3$\sigma$ except
for GRB981226 which is 2$\sigma$. \\
$^{\mathrm e}$ The best available coordinates and error box radius. 
When no source name is available, we report here the 
 equatorial coordinates $\alpha_{J2000}$, $\delta_{J2000}$, when available. \\
$^{\mathrm f}$ No more sensitive follow--up of this source is availiable. 
No decay was observed. No detection of radio or near infrared (NIR) 
afterglow was obtained (\cite{Gorosabel99,Groot97,Wheeler97}) . \\
$^{\mathrm g}$ It is unclear whether this flux was due to the afterglow.
After 1.5 hrs from the first detection the source was no longer visible
\cite{Marshall98}. This would imply a decay index $\beta<-3$. 
Most likely, the \emph{PCA}
source is not related to the GRB \cite{VanParadijs00}. \\
$^{\mathrm h}$ The source was not visible at the beginning 
of the observation campaign. Afterward,
the source flux increased by more than a factor of 3 in about 10000 s. Then
the source started fading. \\
$^{\mathrm i}$ Due to technical problems, the \sax\ observation had a
  duration of only 20 min. The probability of a serendipitous source is
  $\sim 3\times10^{-3}$ \cite{Piro99c}. \\
$^{\mathrm j}$ The source fading was not observed. The probability of
  detecting an unrelated foreground source at this flux level 
 or higher, in the GRB error
  box given by the \sax\ WFCs (see GCN 435) is about 10\%. \\
$^{\mathrm k}$ The source faded to a 2$\sigma$ upper limit of
  $5.0\times10^{-14}$ erg cm$^{-2}$ s$^{-1}$ after 78.86--99.06 hrs from the
  GRB onset. \\
$^{\mathrm l}$ The source was undetected (2$\sigma$ upper limit of
  $1.3\times10^{-13}$ erg cm$^{-2}$ s$^{-1}$) 17 hrs after the main event. \\
$^{\mathrm m}$ Flux in the 1.6--4 keV band. The source was
   undetected (2$\sigma$ upper limit of $3.0\times10^{-14}$~ergs
   cm$^{-2}$ s$^{-1}$) after 20 hrs from the GRB event. This might be a
   field source unrelated to the GRB. \\
$^{\mathrm n}$ Flux in the 0.5-10 keV band.\\
$^{\mathrm o}$ Flux in the 0.2-10 keV band.
\label{t:tab1}
\end{table}

\begin{table}
\caption{X--ray Afterglow Properties}
\renewcommand{\arraystretch}{1.5}
\setlength\tabcolsep{1.0pt}
\scriptsize
\begin{tabular}{lcccccc}
\hline\noalign{\smallskip}
 GRB & Temporal & Photon & n$_H$ /n$_H^G$ & 2--10 keV flux & Redshift & Ref. \\
     & index    & index  &                &  at 10$^5$ s$^{\mathrm a}$& z &  \\
   &$\beta$& I & ($\times10^{21}$ cm$^{-1}$) & (erg cm$^{-2}$s$^{-1}$) & &  \\ 
\noalign{\smallskip}
\hline
\noalign{\smallskip}
 970111 & $<$-1.5 (3$\sigma$) & ... & ... & $<1.0\times10^{-13}$ & ... & \cite{Feroci98a} \\
 970228 & -1.33$^{+0.11}_{-0.13}$ & -2.06$\pm$0.24 & 3.5$_{-2.3}^{+3.3}$ / 1.6 & 
$\sim6.8\times10^{-13}$ &  0.695 & \cite{Costa97,Frontera98b} \\
        & -1.50$^{+0.35}_{-0.23}$ & -2.1$\pm$0.3 & 3.5$_{-2.3}^{+3.3}$ / 1.6 & 
$\sim1.1\times10^{-12}$ & ... & \cite{Frontera98a} \\
 970402 & -1.3--1.6 & -1.7$\pm$0.6 & $<$20 / 2.0 & $\sim4.5\times10^{-14}$ & ... & \cite{Nicastro98a} \\
 970508 & -1.1$\pm$0.1$^{\mathrm b}$  & -1.5$\pm$0.8$^{\mathrm c}$ & 6.0$_{-5.5}^{+13}$ / 0.5 & $3.5\times10^{-13}$$^{\mathrm d}$   & 0.835 & 
 \cite{Piro98b,Piro99a} \\
        &   & -2.2$\pm$0.7$^{\mathrm e}$ &  0.5$_{-0.5}^{+5}$ / 0.5 & $1.5\times10^{-12}$$^{\mathrm f}$ &  & \\
 970828 & -1.44$\pm$0.07 & [-2]$^{\mathrm g}$ & 4.1$_{-1.6}^{+2.1}$ / 0.34$^{\mathrm h}$ & $6.0\times10^{-13}$ &  0.9758 & \cite{Djorgovski01b,Marshall97a,Marshall97b} \\
 971214 & $-0.85^{+0.17}_{-0.15}$ & -1.6$\pm$0.2 & 1.0$_{-1.0}^{+2.3}$ / 0.6 & ... & 3.42 
 & \cite{Dalfiume00} \\
 971227 & -1.12$^{+0.05}_{-0.08}$ & [-2.1] & [0.13] / 0.13 & $\sim1.4\times10^{-13}$ &
  ... &  \cite{Antonelli99a} \\
 980329 & -1.35$\pm$0.03 & -2.4$\pm$0.4 & 10$\pm$4 / 0.9 &  $2.0\times10^{-13}$ &  $<$3.9 
 & \cite{Djorgovski01a,Holland00,Zand98} \\
 980425$^{\mathrm i}$ & -0.16$\pm$0.04 & -2.0$\pm$0.3 & [0.39] / 0.39 & $\sim4.0\times10^{-13}$ &  0.0085 
 & \cite{Pian00} \\
 980519 & -1.83$\pm$0.30 & -2.8$^{+0.5}_{-0.6}$ & 3--20 / 1.73 & $8.0\times10^{-14}$ & ... 
 &  \cite{Nicastro99a} \\
 980613 & -1.19$\pm$0.17 & $>$-2.4 ? & ... & $\sim2.3\times10^{-13}$ & 1.097 
 & \cite{Costa98,Soffitta01} \\
 & -1.05$\pm$0.37$^{\mathrm j }$ & ... & ...  & ...  & ... &  \cite{Boer00,Hjorth02} \\
 980703 & $<$-0.91 ($3\sigma$) & -2.51$\pm$0.32 & 36$_{-13}^{+22}$  / 0.34$^{\mathrm k}$ & $4.5\times10^{-13}$ & 0.966 
 & \cite{Vreeswijk99a} \\
 981226 & -1.31$^{+0.39}_{-0.44}$$^{\mathrm l}$ & -1.92$\pm$0.47 & [0.18] / 0.18 & 
  $\sim2.0\times10^{-13}$ &  ... & \cite{Frontera00a} \\
 990123 & -1.44$\pm$0.07 & -1.86$\pm$0.1$^{\mathrm m }$ & 0.51$_{-0.1}^{+0.7}$ / 0.21 &
 $1.25\times10^{-12}$ &  1.600 & \cite{Heise99,Heise02} \\
 990510 & -1.42$\pm$0.07$^{\mathrm n}$ & -2.03$\pm$0.08 & 2.1$\pm$0.6 / 0.94 & 
 $9.6\times10^{-13}$ & 1.619 & \cite{Kuulkers00a,Pian01} \\
990627 & -1.741$^{+0.63}_{-0.55}$ & -2.8$\pm$0.7 & [0.52] / 0.52 & $<1.6
\times10^{-13}$& \cite{Amati02b} \\
990704 & -0.83$\pm$0.16 & -1.691$^{+0.34}_{-0.60}$ & [0.3] / 0.3 & 
 $\sim3.3\times10^{-13}$ &  ... &  \cite{Feroci01} \\
990705$^{\mathrm o}$ & -1.58$\pm$0.06$^{\mathrm p}$ & ... & ... &
$<1.2\times10^{-13}$ &  0.86 & \cite{Amati00} \\
990806 & -1.2$\pm$0.3 & -2.161$^{+0.50}_{-0.61}$ & [0.35] / 0.35 & 
 $\sim2.0\times10^{-13}$ &  ... & \cite{Montanari01} \\
991014$^{\mathrm q}$ & $<$-0.4 ($3\sigma$) & -1.53$\pm$0.25 & [2.5] / 2.5 & $\sim3.0\times10^{-13}$ &
  ... & \cite{Zand00} \\
991216 & $-1.62\pm 0.07$ & -2.2$\pm$0.2$^{\mathrm r}$ & 3.51$_{-1.5}^{+1.5}$ / 2.1 & 
 $\sim5.0\times10^{-12}$ &  1.02 & \cite{Halpern00,Piro00,Takeshima99c} \\
000115 & $<$-1.0 & ... & ... & $<1.0\times10^{-11}$ & ... & \cite{Marshall00b} \\
000214 & -0.8$\pm$0.5 & -2.0$\pm$0.3$^{\mathrm g}$ & 0.071$_{-0.07}^{+0.75}$ / 0.55 & 
 $\sim5.0\times10^{-14}$$^{\mathrm s}$  & 0.37-- & \cite{Antonelli00} \\ 
   & -1.41$\pm$0.03$^{\mathrm p}$ & ...  & ... & ... & 0.47$^{\mathrm t}$& \\
000926 & -1.89$^{+0.16}_{-0.19}$$^{\mathrm u}$ & -1.85$\pm$0.15$^{\mathrm v}$  
& 4.0$_{-2.5}^{+3.5}$$^{\mathrm w}$ / 0.27 &
 $9.0\times10^{-13}$ &  2.06 & \cite{Harrison01a,Piro01a} \\
 & -1.70$\pm$0.16$^{\mathrm x}$  & -2.23$\pm$0.3$^{\mathrm y}$  &
 [4.0]$^{\mathrm w}$ / 0.27 & ... & ... & \\
001109 & -1.18$\pm$0.05$^{\mathrm p}$  & -2.4$\pm$0.3 & 8.7$\pm$0.4 / 0.42 & 
 $\sim8.0\times10^{-13}$ & ... & \cite{Amati02b} \\
010214  &  -2.1$^{+0.6}_{-1.0}$  & -1.3$ ^{+0.6}_{-0.8}$ & [0.27] / 0.27 &  ... & ... & \cite{Guidorzi03} \\ 
 & for t$>$8 hrs \\
 &   -0.99$^{+0.09}_{-0.04}$ \\
 & for t$<$8 hrs \\
010222 & -1.33$\pm$0.04$^{\mathrm z}$ & -1.97$\pm$0.05 & 1.5$\pm$0.3 / 0.16 & 
 $2.4\times10^{-12}$ & 1.477 & \cite{Zand01} \\
  & for t$>$8 hrs \\
 &   $\sim0.8^{\mathrm p}$ \\
 & for t$<$8 hrs \\
\hline
\end{tabular}
\end{table}

\setcounter{table}{1}
\begin{table} 
\caption{(continued)}
\renewcommand{\arraystretch}{1.5}
\setlength\tabcolsep{7pt}
\scriptsize
%

$^{\mathrm a}$ All upper limits are 3$\sigma$ except for GRB990627 and 
GRB990705 which are 2$\sigma$.\\
$^{\mathrm b}$ from 6$\times$10$^{4}$ s to 5.8$\times$10$^{5}$ s an X--ray 
outburst occurred with integrated energy excess with respect to the 
power--law decay equal to $\sim$5\% that
of the GRB primary event. The outburst was also visible at optical wavelengths 
(see references in \cite{Piro98b}). \\
$^{\mathrm c}$ Index measured before the outburst. During the outburst 
the slope is variable \cite{Piro98b}. \\
$^{\mathrm d}$ Interpolated from the data before and after the outburst. \\ 
$^{\mathrm e}$ Index measured soon after the outburst. \\ 
$^{\mathrm f}$ Interpolated taking into account the outburst data. \\ 
$^{\mathrm g}$ Evidence of an emission feature 
(see Section~\ref{s:features}).\\ 
$^{\mathrm h}$ n$_H$ measured in the latest observation period. \\ 
$^{\mathrm i}$ Assuming that the source S1 is coincident with 
SN1998bw position (see Section~\ref{s:time}). \\ 
$^{\mathrm j}$ Result reported by \cite{Boer00}. \\ 
$^{\mathrm k}$ n$_H$ value corrected for redshift. \\ 
$^{\mathrm l}$ Temporal index during the fading phase of the source. 
See note h of Table 1. \\ 
$^{\mathrm m}$ The afterglow is visible up to 60 keV with a 
power-law spectrum. \\ 
$^{\mathrm n}$ Comparing the prompt emission with the X-ray afterglow 
 emission, Pian et al. \cite{Pian01} derive the presence of a break 
in the afterglow light curve which tends, at early and late epochs, 
to a power--law of  index $\beta\sim-1$ and $\beta\sim-2$,
 respectively, with  a break time of $\sim$ 0.5 days, 
compatible with that determined from the optical light curve 
\cite{Harrison99,Israel99,Stanek99}. \\
$^{\mathrm o}$ The field is highly contaminated by stray radiation. \\
$^{\mathrm p}$ Derived by comparing the prompt emission and late 
afterglow data.\\
$^{\mathrm q}$ The shortest event followed by \sax\ 
(3.29 s at 40--700 keV). \\
$^{\mathrm r}$ Also evidence of two emission lines (see Table 3).  \\
$^{\mathrm s}$ Extrapolated from the measured light curve. \\
$^{\mathrm t}$ Based on X--ray emission feature interpretation (see text).\\
$^{\mathrm u}$ \sax\ plus \emph{CHANDRA} data. 
Evidence of a discrepancy between optical and X--ray decline rates
 \cite{Piro01a}. \\
$^{\mathrm v}$  \sax\ plus \emph{CHANDRA} data 2--3 days 
after the main event. \\
$^{\mathrm w}$ Corrected for redshift \cite{Piro01a}. This n$_H^z$ value
was added to the Galactic column density n$_H^G$. According to 
Harrison et al. \cite{Harrison01a}, the n$_H^z$  is not needed by the \emph{CHANDRA} data. \\
$^{\mathrm x}$ From \emph{CHANDRA} data only. \\
$^{\mathrm y}$ From \emph{CHANDRA} data 13 days after the GRB.\\
$^{\mathrm z}$ The extrapolation of the power--law light
curve, derived from the \sax\ data, is above the X--ray flux measured 9
days after the burst with \emph{CHANDRA} \cite{Harrison01b,Zand01} 
(see text). \\
\label{t:tab2}
\end{table}

\begin{table}[t]
\caption{Properties of the Detected X--ray Emission Features}
\label{t:tab3}
\vspace{0.4cm}
\begin{center}
\begin{tabular}{lllll}
\hline\noalign{\smallskip}
Quantity  & GRB970508 & GRB970828 & GRB991216 & GRB000214      \\
\noalign{\smallskip}
\hline
\noalign{\smallskip}
$E_{l}$ (keV) & $3.4\pm 0.3$ & $5.04^{+0.23}_{-0.31}$ & $3.49\pm 0.06$
& $4.7\pm 0.2$ \\
  &  &  & $4.4 \pm 0.5$ &  \\
$FWHM_{l}$ (keV) & $< 0.5$ & $0.73^{+0.89}_{-0.73}$ & $0.54 \pm 0.16$ &
$<0.4$ \\
  &  &  & $> 2.35$ &  \\
$F_{l}^{\mathrm a}$ & $3\pm 1$ & $1.5 \pm 0.8$ & $1.7 \pm 0.5$ & 
$0.67 \pm 0.22$ \\
($10^{-13}$erg cm$^{-2}$ s$^{-1}$) &  &  &  $2.7 \pm 1.4$ &  \\
$EW_l$ (keV) & $\sim 1.5$ & $\sim 1.8$ & $0.5\pm 0.013$ & $\sim 2.1$\\ 
    &  &  & $\sim 0.8$ &  \\
$F_{\nu}^{\mathrm b}$  & $\sim 7$ & $\sim 0.4$  & $\sim 2.3$ & $\sim 0.3$   \\
($10^{-12}$ erg cm$^{-2}$ s$^{-1}$) &  &  &  &       \\ 
Redshift $z$ & 0.835  & 0.9578   & 1.02 & 0.37--0.47 $^{\mathrm c}$    \\ 
$E_{l}^0$ (keV)$^{\mathrm d}$ &  $6.24 \pm 0.55$ & $9.86^{+0.45}_{-0.61}$ &
  $7.05\pm 0.12$ &  6.4--6.97 $^{\mathrm b}$ \\
        &  &  & $8.9 \pm 1.0$ &  \\
$L_{l}^{\mathrm e}$ & $6.7\pm2.7$ & $4.8\pm2.5$ & $5.7\pm 1.6$ &
 $0.45\pm 0.15$ $^{\mathrm f}$  \\
($10^{44}$ erg s$^{-1}$)   &  &  & $9.0 \pm 4.7$ &  \\
hrs from GRB$^{\mathrm g}$  & 6--16 & 32--38 & 37--40 & 12--41 \\ 
Mission & \emph{BeppoSAX} & \emph{ASCA} & \emph{CHANDRA} & \emph{BeppoSAX} \\
Instrument  &  \emph{NFI}     &  \emph{SIS+GIS} & \emph{Gratings+ACIS-S} &
  \emph{NFI}  \\
References & \cite{Piro99a} & \cite{Yoshida99} & \cite{Piro00} &
\cite{Antonelli00} \\
\noalign{\smallskip}
\hline
\end{tabular}
\end{center}
$^{\mathrm a}$ Line flux.\\
$^{\mathrm b}$ Continuum flux in the 2--10 keV band.\\
$^{\mathrm c}$ Depends on the Fe ionization state (Fe K X--ray
fluorescence assumed).\\
$^{\mathrm d}$ Energy line corrected for cosmological redshift. \\
$^{\mathrm e}$ Derived assuming $H_0=65$km/s Mpc and $q_0=0.5$. \\
$^{\mathrm f}$ Assuming  $z=0.47$. \\
$^{\mathrm g}$ Time interval during which the line was visible. \\

\end{table}

\end{document}